\documentclass[12pt, eqsecnum]{revtex4}
\begin{document}
\title{The model for self-dual chiral bosons as a Hodge theory}

%\vspace{1.3cm}
\author{ Sudhaker Upadhyay\footnote {e-mail address: sudhakerupadhyay@gmail.com}}
\author{ Bhabani Prasad Mandal\footnote{e-mail address:
\ \ bhabani.mandal@gmail.com,\ \ bhabani@bhu.ac.in  }}

\affiliation { Department of Physics,\\
Banaras Hindu University,\\
Varanasi-221005, INDIA. \\
}
\begin{abstract}
We consider (1+1) dimensional theory for a single self-dual 
chiral boson as classical model for gauge theory.
Using Batalin-Fradkin-Vilkovisky (BFV) technique the nilpotent BRST and anti BRST symmetry 
transformations for this
theory have been studied. In this model 
other forms of nilpotent symmetry transformations like co-BRST and anti co-BRST  
which leave the
gauge-fixing part of the action invariant, are also explored. We show that the nilpotent 
charges for these symmetry 
transformations satisfy the  algebra
of de Rham cohomological operators in differential geometry. 
The Hodge decomposition theorem on compact manifold is also studied in the context of 
conserved charges.
\end{abstract}
\maketitle
Keywords: {Chiral Boson; BFV approach; BRST transformation;  
 Hodge theory;\\ Cohomological 
algebra. }\\

\section{Introduction}
\label{intro}
The quantization of classically formulated model and the investigation of their properties 
is of particular importance in the study of quantum field theory, especially for the case
in which the theory possesses constraints. The Dirac quantization \cite{dir} of these 
theories requires that the Dirac brackets to be calculated with respect to the field 
variables 
and the constraints of the theory. Another way of approaching the problem, which is very
different from the Dirac's method, is the BFV quantization 
\cite{hen,hen1}. It is a powerful technique to study the 
BRST quantization of constrained systems. The main features of BFV approach are, (I)
it does not require closure off-shell of the gauge algebra and therefore does not need an 
auxiliary field, (II) this formalism relies on BRST transformation which are independent 
of gauge fixing condition and (III) it is also applicable to the first order Lagrangian 
and hence is more general than the strict Lagrangian approach.

In the formulation of the theories with first class constraints one requires that
 the physical subspace of total Hilbert space of states contains only those states that 
are annihilated by the nilpotent and conserved BRST charge $Q_b$ i.e. 
$Q_b\left|phys\right>=0$ \cite{dir,sund}.
The nilpotency  of the BRST charge ($Q^2_b = 0$)
and the physicality criteria ($Q_b\left|phys \right>= 0$) are the two essential ingredients
 of BRST quantization. In the language of differential geometry  defined on compact,
 orientable
 Riemannian manifold, the cohomological aspects of BRST charge is realized in a simple, 
elegant manner. The nilpotent BRST charge is connected with  exterior derivative 
(de Rham cohomological operator $d = dx^\mu\partial_\mu,$ with $ d^2 =
0$)\cite{egu,nis,nis1,kala,hol,hol1,arn,hari,sr,sr1}. It has been found that the co-BRST 
transformation which is also the symmetry of the action and leaves the gauge fixing 
part of the action invariant. The conserved charge corresponding the co-BRST 
transformation is shown to be analogue of co-exterior derivative 
( $\delta = \pm \ast d \ast,$ with $ \delta^2 =
0$)\cite{hari}.

In this work we consider the gauge non-invariant model
 for single self-dual 
chiral boson \cite{kk,pp,pp1,fj,vo}. With addition of the Wess-Zumino term this model 
is treated as a classical system for first class constraint in the context of BFV 
formulation. Such a model is very useful in the study of certain string theoretic 
model \cite{ms}
and plays a crucial role in the studies of quantum Hall effect \cite{wen}. 
Along with nilpotent BRST symmetry,  anti BRST (where the role of ghost and antighost fields 
are changed with some changes in coefficients), co-BRST and anti co-BRST transformations 
have been investigated in this framework.
 The generators of all these continuous 
symmetry transformation
are shown to obey the algebra of de Rham cohomological operators of differential geometry.
Hodge decomposition theorem in quantum Hilbert space of states is also discussed. Thus
we show that the classical theory for self-dual chiral boson is field theoretic model for 
Hodge theory.

The present model for self-dual chiral boson in (1+1) dimension reduces to an effective 
model in (0+1) dimension due to self-duality condition. As a consequence, the analogue of 
the Hodge duality ($\ast $) operation cannot be defined for this effective model.

We start with the brief introduction about the classical model in Sec. II.
 In Sec. III the BFV formulation for the same model 
has been discussed and BRST symmetry for such model has been constructed. 
Sec. IV is devoted to show the other nilpotent symmetry transformations for the same system. 
The connection between algebra satisfied by nilpotent charges and de 
Rham cohomological operators of differential geometry is shown in Sec V. We conclude
our results in Sec. VI.
\section{Gauge invariant theory for self-dual chiral boson: Preliminary idea}
\label{sec:1}  We start with the gauge non-invariant model \cite{pp} in (1+1) dimension for
 single self-dual chiral boson. The Lagrangian density for such a theory is given as 
\begin{equation} 
{\cal L}=\frac{1}{2}\dot\varphi^2 -\frac{1}{2}{\varphi'}^2 +\lambda (\dot\varphi -\varphi' ),
\label{chi}
\end{equation}  
where over dot and prime 
denote time and space derivatives respectively and $\lambda$ is Lagrange multiplier.
The field $\varphi$ satisfies the self-duality condition $\dot\varphi =\varphi'$ 
in this case.
We choose the Lorentz metric $g^{\mu\nu}=(1,-1)$ where $\mu, \nu =0, 1$. The associated 
momenta for the field  $\varphi$ and Lagrange multiplier are calculated as
\begin{eqnarray}
\pi_\varphi &=&\frac{\partial {\cal L}}{\partial \dot \varphi}=\dot\varphi +
\lambda\nonumber\\
\pi_\lambda &=&\frac{\partial {\cal L}}{\partial \dot \lambda}=0,
\end{eqnarray}
which shows that the model has following primary constraint
\begin{equation}
\Omega_1\equiv \pi_\lambda \approx 0.
\end{equation}
The expression for Hamiltonian density corresponding to above Lagrangian density ${\cal L}$ 
is
\begin{eqnarray}
{\cal H} &=&\pi_\varphi\dot\phi +\pi_\lambda\dot\lambda -{\cal L}\nonumber\\
&=& \frac{1}{2}(\pi_\varphi -\lambda )^2 +\frac{1}{2}\varphi'^2 +\lambda \varphi'.
\end{eqnarray}
Further we can write the total Hamiltonian density corresponding to ${\cal L}$ by 
introducing  Lagrange multiplier field  $\omega$ for the primary constraint $\Omega_1$
 as
\begin{eqnarray}
{\cal H}_T = \frac{1}{2}(\pi_\varphi -\lambda )^2 +\frac{1}{2}\varphi'^2 +\lambda \varphi' +
\omega\pi_\lambda.
\end{eqnarray}
Following the Dirac prescription \cite{dir}, we obtain the secondary constraint in this case
as
\begin{equation}
\Omega_2 \equiv \dot \pi_\lambda =\{ \pi_\lambda, {\cal H}\}=\pi_\varphi -
\lambda-\varphi'\approx 0.
\end{equation}  
 The Poison bracket for primary and secondary constraint is nonvanishing,
$\{ \Omega_1, \Omega_2\} \not=0$.
 This implies the constraints $\Omega_1$ and $\Omega_2$ are of second class,
which is an essential feature of gauge variant theory (model).

This model is quantized by establishing the following commutation relation \cite{pp}
\begin{eqnarray}
[\varphi(x), \pi_\varphi (y)]& =&[\varphi(x), \lambda(y) ]=-i\delta (x-y)\\
2[\lambda(x), \pi_\varphi(y)] &=&[\lambda(x), \lambda(y) ]= 2i\delta' (x-y),
\end{eqnarray}
and rest of the commutators vanish.
\subsection{Wess-Zumino term and Hamiltonian formulation}
\label{sec:2}
 To construct a gauge invariant theory corresponding to this gauge non-invariant model
for chiral bosons, one generally introduces the Wess-Zumino term in the Lagrangian density
${\cal L}$.
For this purpose we enlarge the Hilbert space of the theory by introducing a new 
quantum field 
$\vartheta$, called as Wess-Zumino field, through the redefinition of
 fields $\varphi$ and $\lambda$ in the original Lagrangian 
density ${\cal L}$ [Eq. (\ref{chi})] as follows \cite{wz}
\begin{equation}
\varphi\rightarrow \varphi -\vartheta, \ \ \lambda\rightarrow \lambda +\dot\vartheta.
\end{equation}
With these  redefinition of fields the modified Lagrangian density becomes
\begin{eqnarray}
{\cal L}^I&=& {\cal L} -\frac{1}{2}\dot\vartheta^2 -\frac{1}{2}{\vartheta'}^2 +
\varphi'\vartheta' +\dot\vartheta\vartheta' -\dot\vartheta\varphi' 
-\lambda (\dot\vartheta -\vartheta')\nonumber\\
&=&{\cal L} +{\cal L}^{WZ},
\end{eqnarray}
where,
\begin{equation}
{\cal L}^{WZ}=-\frac{1}{2}\dot\vartheta^2 -\frac{1}{2}{\vartheta'}^2  +
\varphi'\vartheta' +\dot\vartheta\vartheta' -\dot\vartheta\varphi' 
-\lambda (\dot\vartheta -\vartheta'),
\end{equation}
is the
Wess-Zumino part of the  ${\cal L}^I$.
It is easy to check that the above Lagrangian density is invariant under 
time-dependent chiral gauge transformation:
\begin{eqnarray}
\delta\varphi &=&\mu (x, t),\ \ \delta\vartheta =\mu (x, t),\ \ \delta\lambda =-\dot \mu (x, 
t)
\nonumber\\
\delta\pi_\varphi &=&0,\ \ \delta\pi_\vartheta =0,\ \ \delta p_\lambda =0,
\end{eqnarray} 
where $\mu(x, t)$ is an arbitrary function of the space and time.
The canonical momenta for this gauge-invariant theory are calculated as
\begin{eqnarray}
\pi_\lambda &= &\frac{\partial{\cal L}^I}{\partial\dot\lambda}=0,\ \
\pi_\varphi = \frac{\partial{\cal L}^I}{\partial\dot\varphi}=\dot\varphi +\lambda\nonumber\\
\pi_\vartheta & =&\frac{\partial{\cal L}^I}{\partial\dot\vartheta}=-\dot\vartheta 
-\varphi' +\vartheta' -\lambda.
\end{eqnarray}
This implies the theory ${\cal L}^I$ possesses a primary constraint
\begin{equation}
 \psi_1\equiv \pi_\lambda \approx 0.
\end{equation}
The Hamiltonian density corresponding to ${\cal L}^I$ is then given by
\begin{equation}
{\cal H}^I= \pi_\varphi\dot\varphi +\pi_\vartheta\dot\vartheta +\pi_\lambda\dot\lambda
-{\cal L}^I.
\end{equation}
The total Hamiltonian density after the introduction of a Lagrange multiplier field $u$ 
corresponding to the primary constraint $\Psi_1$ becomes
\begin{equation}
{\cal H}_T^I=\frac{1}{2}\pi_\varphi^2 -\frac{1}{2}\pi_\vartheta^2 +\pi_\vartheta\vartheta'-
\pi_\vartheta\varphi' -\lambda\pi_\varphi -\lambda\pi_\vartheta +u\pi_\lambda.\label{ham}
\end{equation}
Following the Dirac's method of constraint analysis we obtain secondary
constraint 
\begin{equation}
\psi_2\equiv (\pi_\varphi +\pi_\vartheta)\approx 0.
\end{equation}
In Dirac's quantization procedure \cite{dir}, we have to change the first class 
constraints of the theory into second class constraints. To achieve this we impose some
additional constraints on the system in the form of gauge fixing conditions $\partial_\mu
\vartheta=0$ ($\partial_0\vartheta=\dot\vartheta=0$ and $-\partial_1\vartheta 
=-\vartheta'=0$)
\cite{nk}.
With the above choice of gauge fixing conditions the extra constraint of the theory are
\begin{eqnarray}
\xi_1\equiv  -\vartheta'\approx 0, \ \ \xi_2 \equiv (\pi_\vartheta -\vartheta' +\varphi' +
\lambda)\approx 0.\label{cons}
\end{eqnarray}
Now, the total set of constrains after gauge fixing are
\begin{eqnarray}
\chi_1 &=&\psi_1 \equiv \pi_\lambda\approx 0, \ \ \chi_2 = \psi_2 \equiv (\pi_\varphi 
+\pi_\vartheta)\approx 0\nonumber\\
\chi_3 &=&\xi_1\equiv  -\vartheta'\approx 0, \ \ \chi_4 =\xi_2 \equiv (\pi_\vartheta -
\vartheta' 
+\varphi' +\lambda)\nonumber\\
\approx 0.
\end{eqnarray}
Using Dirac's quantization procedure \cite{dir}, the nonvanishing
commutators of gauge invariant theory are obtained as
\begin{eqnarray}
[\varphi(x), \pi_\varphi (y)]& =&[\varphi(x), \lambda(y) ]=-i\delta (x-y)\label{brac1}\\
2[\lambda(x), \pi_\varphi(y)] &=&[\lambda(x), \lambda(y) ]=+2i\delta' (x-y)\\
\ [ \vartheta (x), \pi_\vartheta(y) ] &=& 2[\varphi(x), \pi_\vartheta(y) ]=-2i\delta (x-y)\\
\ [\lambda(x), \pi_\vartheta(y)] &=&-i\delta' (x-y)\label{brac4}.
\end{eqnarray}
We end up the section with conclusion that the above relations (\ref{brac1})-(\ref{brac4})
together with ${\cal H}^I_T$ [Eq. (\ref{ham})],
reproduce the same quantum system describe by ${\cal L}$ under the gauge condition 
(\ref{cons}). This is similar to the quantization
of a gauge invariant chiral Schwinger model (with an appropriate Wess-Zumino term)\cite{nk}.
\section{BFV formulation for model of self-dual chiral boson }
In the BFV approach of this model, one needs to introduce a pair of canonically conjugate 
ghosts $(c, p)$ with
ghost number 1 and $-1$ respectively, for the first class constraint, $\pi_\lambda=0$ and 
another pair of
 ghosts $( \bar c, \bar p)$ with ghost number $-1$ and 1 respectively, for other 
constraint,
$(\pi_\varphi +\pi_\vartheta) = 0 $. The effective action for (1+1) dimensional 
theory for a single self-dual 
chiral boson in this extended 
phase space then becomes
\begin{eqnarray}
S_{eff}& =& \int d^2x  \left[ \pi_\varphi\dot\varphi +\pi_\vartheta\dot\vartheta +p_u\dot u
-\pi_\lambda\dot\lambda
-\frac{1}{2}\pi_\varphi^2 +\frac{1}{2}\pi_\vartheta^2 \right.\nonumber\\
&-&\pi_\vartheta(\vartheta'-\varphi')+\left.\dot cp +\dot{\bar c}\bar p
 - \{Q_b, \Psi\}\right],
\label{seff}
\end{eqnarray}
where $Q_b$ is the BRST charge and $\Psi$ is the gauge fixed fermion.

The generating functional, for any gauge invariant effective theory having $\Psi$ as a
 gauge fixed fermion, is defined as
\begin{equation}
Z_\Psi = \int {\cal{D}}\phi \exp \left[i \int d^2x \ S_{eff} \right],
\end{equation}
where $\phi$ is generic notation for all the dynamical field involved in the effective 
theory.
The BRST symmetry generator for this theory is written as
\begin{equation}
Q_b=  ic(\pi_\varphi +\pi_\vartheta )-i\bar p \pi_\lambda .\label{chr}
\end{equation}
The canonical brackets are defined for all dynamical variables as
\begin{eqnarray}
[\vartheta, \pi_\vartheta ]&=&-i,\ \ [\varphi, \pi_\varphi ]=-i, \ \ [\lambda, \pi_\lambda] 
= -i,
\nonumber\\
\ [u, p_u]&=&-i,\ \ \{ c,\dot{\bar c}\} =i, \ \ \{ \bar c,\dot{ c}\} = -i, \label{brac}
\end{eqnarray}
and rest of the brackets are zero. The nilpotent BRST transformation, using the relation 
$s_b\phi = -[\phi,Q_b]_\pm $ ($\pm $ sign 
represents the bosonic and fermionic nature of fields $\phi$),
can explicitly be written as
\begin{eqnarray}
s_b \varphi &=& -c,\ \ \ s_b\lambda =\bar p,\ \ \ s_b\bar p=0, \ \ \ s_b\vartheta =-c
\nonumber\\
s_b \pi_\varphi &=& 0,\ \ \ s_b u =0, \ \ \ s_b \pi_\vartheta =0,\ \ s_b p=(
\pi_\varphi +\pi_\vartheta )\nonumber\\
s_b \bar c &=& \pi_\lambda,\ \ \ s_b \pi_\lambda =0, \ \ \ s_b c =0,\ \ s_b p_u =0.
\end{eqnarray}
In BFV formulation the generating functional is independent of the
gauge fixed fermion \cite{hen1,wei}, hence we have liberty to choose it in the convenient
 way as
\begin{equation}
\Psi = {{p}}\lambda +{\bar{{c}}} \left (\vartheta +\varphi +\frac{\xi}{2}
\pi_\lambda\right ),
\end{equation}
where $\xi$ is arbitrary gauge parameter.

Putting the value of $\Psi$ in Eq. (\ref{seff}) and using Eq. (\ref{chr}), we get
\begin{eqnarray}
S_{eff}& =& \int d^2x \left[ \pi_\varphi\dot\varphi +\pi_\vartheta\dot\vartheta +p_u\dot u
-\pi_\lambda\dot\lambda
-\frac{1}{2}\pi_\varphi^2 +\frac{1}{2}\pi_\vartheta^2 \right.\nonumber\\
&-&\left.\pi_\vartheta(\vartheta'-\varphi') 
+\dot cp +\dot{\bar c}\bar p +\lambda (\pi_\varphi +\pi_\vartheta )
+2 c\bar c -\bar pp\right.\nonumber\\
&+&\left.\pi_\lambda \left(\vartheta +\varphi +\frac{\xi}{2}\pi_\lambda\right )
\right]. \label{effact}
\end{eqnarray}
The generating functional for this effective theory  can then be expressed as 
\begin{eqnarray}
Z_\Psi &=& \int {\cal{D}}\phi \exp \left[i \int d^2x \left\{ \pi_\varphi\dot\varphi +
\pi_\vartheta\dot\vartheta +p_u\dot u
-\pi_\lambda\dot\lambda\right.\right.\nonumber\\
&-&\left.\left.\frac{1}{2}\pi_\varphi^2 
+\frac{1}{2}\pi_\vartheta^2 
-\pi_\vartheta(\vartheta'-\varphi') 
+\dot cp +\dot{\bar c}\bar p +\lambda (\pi_\varphi +\pi_\vartheta )\right.\right.\nonumber\\
&+&\left.\left.2 c\bar c 
-\bar pp+\pi_\lambda \left(\vartheta +\varphi +\frac{\xi}{2}\pi_\lambda\right )
\right\}
\right]. 
\end{eqnarray}
Performing the integration over $p$ and $\bar p$ in the above functional integration 
we further obtain
\begin{eqnarray}
Z_\Psi &=& \int {\cal{D}}\phi' \exp \left[i \int d^2x\left\{ \pi_\varphi\dot\varphi +
\pi_\vartheta\dot\vartheta +p_u\dot u
-\pi_\lambda\dot\lambda\right.\right.\nonumber\\
&-&\left.\left.\frac{1}{2}\pi_\varphi^2 
+\frac{1}{2}\pi_\vartheta^2 -
\pi_\vartheta(\vartheta'-\varphi')
+\dot{\bar c}\dot c  +\lambda (\pi_\varphi +\pi_\vartheta )\right.\right.\nonumber\\
&+&\left.\left.2 c\bar c 
+\pi_\lambda \left(\vartheta +\varphi +\frac{\xi}{2}\pi_\lambda\right )\right\}
\right],
\end{eqnarray}
where ${\cal D}\phi'$ is the path integral measure for effective theory when integrations
 over 
fields $p$ and $\bar p$ are carried out. 
Taking the arbitrary gauge parameter $\xi =1$ and performing the integration over 
$\pi_\lambda$, we obtain an 
effective generating functional as
\begin{eqnarray}
Z_\Psi &=& \int {\cal{D}}\phi'' \exp \left[i \int d^2x \left\{ \pi_\varphi\dot\varphi +
\pi_\vartheta\dot\vartheta +p_u\dot u
-\frac{1}{2}\pi_\varphi^2  \right.\right.\nonumber\\
&+&\left.\left.\frac{1}{2}\pi_\vartheta^2 +
\pi_\vartheta (\varphi'-\vartheta' +\lambda )
+\dot{\bar c}\dot c +\pi_\varphi\lambda  
-2\bar cc \right.\right.\nonumber\\
 &-&\left.\left.\frac{ \left (\dot\lambda -\vartheta -\varphi \right )^2}{2}\right\}
\right], \label{zfun}
\end{eqnarray}
where ${\cal D}\phi''$ denotes the measure corresponding to all the dynamical
 variable involved in this effective action. 
The expression for effective action in the above equation is exactly 
same as the BRST invariant effective action in Ref. \cite{pb}. 
The BRST symmetry transformation for this effective theory is
\begin{eqnarray}
s_b \varphi &=& -c,\ \ \ s_b\lambda =\dot c, \ \ \ s_b\vartheta =-c,
\nonumber\\
s_b \pi_\varphi &=& 0,\ \ \ s_b u =0, \ \ \ s_b \pi_\vartheta =0,\nonumber\\
s_b \bar c &=&-(\dot\lambda -\vartheta -\varphi ), \ \ \ s_b c =0,
\nonumber\\ 
s_b p_u& =&0,\label{brs}
\end{eqnarray}
which is {\it on-shell} nilpotent. Antighost equation of motion (i.e. $\ddot{ c}+2c=0$)
 is required
to show the nilpotency.
\section{Nilpotent symmetries: many guises}
In this section we study other different nilpotent symmetries of the models with self dual 
chiral boson, which have different characteristics 
and play important role in the study of gauge field theories.
\subsection{ Off-shell  BRST and anti-BRST Symmetry }
We incorporate the Nakanishi-Lautrup type auxiliary field $B$  to linearize the gauge fixing 
part of the effective action. The first order effective Lagrangian density is then given by
 \begin{eqnarray}
{\cal  L}_{eff}&=& \pi_\varphi\dot\varphi +\pi_\vartheta\dot\vartheta +p_u\dot u-\frac{1}{2}
\pi_\varphi^2
+\frac{1}{2}\pi_\vartheta^2 \nonumber\\
&+&\pi_\vartheta (\varphi'-\vartheta' +\lambda)
+ \pi_\varphi\lambda +\frac{1}{2}B^2 +B(\dot \lambda -\varphi -\vartheta )\nonumber\\
&+&\dot{\bar c}\dot c
-2\bar c c.\label{lag}
\end{eqnarray}
One can trivially show that this effective theory is invariant under the following {\it 
off-shell} nilpotent 
BRST transformation,
\begin{eqnarray}
s_{b}\varphi &=& - c,\ \ \ s_{b}\lambda =\dot { c
}, \ \ \ s_{b}\vartheta = -c\nonumber\\
s_{b} \pi_\varphi &=& 0,\ \ \ s_{b} u =0, \ \ \ s_{b} \pi_\vartheta =0
\nonumber\\
s_{b}\bar c &=& B,\ \ \ s_{b} B =0, \ \ \ s_{b}  c =0,\ \ s_{b} p_u 
=0.\label{b}
\end{eqnarray}
The corresponding anti BRST symmetry transformation 
of this theory can be written as
\begin{eqnarray}
s_{ab}\varphi &=& -\bar c,\ \ \ s_{ab}\lambda =\dot {\bar c
}, \ \ \ s_{ab}\vartheta = -\bar c\nonumber\\
s_{ab} \pi_\varphi &=& 0,\ \ \ s_{ab} u =0, \ \ \ s_{ab} \pi_\vartheta =0
\nonumber\\
s_{ab}c &=& -B,\ \ \ s_{ab} B =0, \ \ \ s_{ab} \bar c =0,\ \ s_{ab} p_u 
=0.\label{ab}
\end{eqnarray}
The conserved  BRST and anti BRST charge
$Q_{b}$ and $Q_{ab}$ respectively, that are the generators of the 
above  BRST and anti BRST symmetry transformations, are
\begin{equation}
Q_b=  i(\pi_\varphi +\pi_\vartheta )c -i\pi_\lambda \dot c 
\end{equation}
and
\begin{equation}
Q_{ab}=  i(\pi_\varphi +\pi_\vartheta )\bar c -i\pi_\lambda \dot{\bar c}.
\end{equation}
Further by using the following equation of motion
\begin{eqnarray}
B+\dot \pi_\varphi =0,\ \ B+\dot \pi_\vartheta =0,\ \ \dot\varphi-\pi_\varphi +\lambda =0
\nonumber\\
\dot\vartheta +\pi_\vartheta +\varphi' -\vartheta' +\lambda =0,\ \ \dot B=\pi_\varphi +
\pi_\vartheta \nonumber\\
\dot u=0,\ \ \dot p_u =0, \ \ \ddot{\bar c}+2\bar c=0, \ \ \ddot{ c}+2 c=0\nonumber\\
B+\dot \lambda -\varphi -\vartheta =0,\label{eom}
\end{eqnarray}
it can be shown that these charges are constant of motion i.e. $\dot Q_b=0,\dot Q_{ab}=0$,
which satisfy following relation,
\begin{equation}
Q_b^2=0,\ Q_{ab}^2=0,\ Q_b Q_{ab}+Q_{ab}Q_b=0.
\end{equation}
To arrive in these relations, we have used the canonical brackets 
[Eq. (\ref{brac})] of the fields and the definition
of canonical momenta, 
\begin{equation}
\pi_\lambda =B,\ \ \pi_{\bar c} =\dot c,\ \ \pi_c =-\dot {\bar c},\ \ \pi_u =p_u.
\end{equation} 
We come to the end of this section with the remark that the condition for the physical states
$Q_{b}\left|phys\right>=0$ and $Q_{ab}\left|phys\right>=0$ leads to the requirement that
\begin{equation} 
(\pi_\varphi +\pi_\vartheta)\left|phys\right>
=0
\end{equation} and 
\begin{equation}
\pi_\lambda\left|phys\right>=0.
 \end{equation}
This implies
that the operator form of the first class constraint $\pi_\lambda\approx 0$ and 
$(\pi_\varphi +\pi_\vartheta)\approx 0$ annihilate the physical state of the theory.
Thus the physicality criteria is consistent with the Dirac's method \cite{sund} of 
quantization.
\subsection{ Co-BRST and anti co-BRST symmetries} 
The gauge fixing term has its origin in the co-exterior derivative $\delta =\pm \ast d\ast $,
where $\ast $ represents the Hodge duality operator. The $\pm$
signs is dictated by the dimensionality of the manifold \cite{egu}.
In this subsection we investigate the nilpotent  co-BRST and anti co-BRST (so-called dual 
and anti dual-BRST respectively)
transformation which are also the symmetry of the effective action. Further these
transformations leave the gauge-fixing term of the action invariant independently and the 
kinetic energy term (which remains invariant under BRST and anti-BRST transformations) 
transforms under it to compensate for the transformation of the ghost terms. Therefore 
it is appropriate to call these transformations as dual 
and anti dual-BRST transformation.

The nilpotent co-BRST transformation ($ s^2_{d}=0$) and anti co-BRST transformation ($ 
s^2_{ad}=0$) which are absolutely
 anticommuting ($ s_d s_{ad}+s_{ad}s_d =0$) are                 
\begin{eqnarray}
s_{d}\varphi &=& -\frac{1}{2}\ \dot{\bar c},\ \ \ s_{d}\lambda =-\bar c
, \ \ \ s_{d}\vartheta = -\frac{1}{2} \ \dot{\bar c}\nonumber\\
s_{d} \pi_\varphi &=& 0,\ \ \ s_{d} u =0, \ \ \ s_{d} \pi_\vartheta =0,\ \ s_{d} p_u 
=0
\nonumber\\
s_{d}c &=& \frac{1}{2}(\pi_\varphi +\pi_\vartheta ),\ \ \ s_{d} B =0, \ \ \ s_{d} \bar c =0.
\label{d}
\end{eqnarray}
\begin{eqnarray}
s_{ad}\varphi &=& -\frac{1}{2}\ \dot{c},\ \ \ s_{ad}\lambda =-c
, \ \ \ s_{ad}\vartheta = -\frac{1}{2}\dot{ c}\nonumber\\
s_{ad} \pi_\varphi &=& 0,\ \ \ s_{ad} u =0, \ \ \ s_{ad} \pi_\vartheta =0,\ \ s_{ad} p_u 
=0
\nonumber\\
s_{ad}\bar c &=& -\frac{1}{2}(\pi_\varphi +\pi_\vartheta ),\ \ \ s_{ad} B =0, \ \ \ s_{ad} c 
=0.
\label{ad}
\end{eqnarray}
The conserved charges for the above symmetries (using Noether's theorem) are
\begin{equation}
Q_d = i\frac{1}{2}(\pi_\varphi +\pi_\vartheta )\dot{\bar c}+i\pi_\lambda\bar c
\end{equation}
and
 \begin{equation}
Q_{ad}=i\frac{1}{2}(\pi_\varphi +\pi_\vartheta )\dot{ c}+i\pi_\lambda c.
\end{equation}
$Q_d$ and $Q_{ad}$ generate the symmetry  transformations in Eqs. (\ref{d}) and (\ref{ad}) 
respectively.
It is easy to verify the following relations 
\begin{eqnarray}
s_dQ_d&=&-\{Q_d,Q_d\}=0\nonumber\\
s_{ad}Q_{ad}&=&-\{Q_{ad},Q_{ad}\}=0\nonumber\\
s_dQ_{ad}&=&-\{Q_{ad},Q_d\}=0\nonumber\\
s_{ad}Q_{d}&=&-\{Q_{d},Q_{ad}\}=0
\end{eqnarray}
which reflect the nilpotency and anticommutativity property of $s_{d}$ and 
$s_{ad}$ (i.e.
$ s^2_{d}=0, s^2_{ad}=0$ and $ s_d s_{ad}+s_{ad}s_d =0$).
\subsection{Bosonic symmetry}
In this subsection we construct the bosonic symmetry out of these nilpotent BRST symmetries 
of the theory.
The  BRST ($s_{b}$), anti BRST ($s_{ab}$), co-BRST ($s_{d}$) and anti co-BRST ($s_{ad}$) 
symmetry operators satisfy 
the following algebra 
\begin{equation}
\{ s_d, s_{ad}\} =0,\ \ \{ s_b, s_{ab}\} =0
\end{equation}
\begin{equation}
\{ s_b, s_{ad}\} =0,\ \ \{ s_d, s_{ab}\} =0,
\end{equation}
\begin{equation}
\{ s_b, s_d\} \equiv s_w,\ \ \ \{ s_{ab}, s_{ad}\} \equiv s_{\bar w}.\label{bos}
\end{equation}
The anticommutator in Eq. (\ref{bos}) define the bosonic symmetry of the system.
Under this bosonic symmetry transformation the field variables transform  as
\begin{eqnarray}
s_{w}\varphi &=& -\frac{1}{2}(\dot B +\pi_\varphi +\pi_\vartheta ),
\ \ s_w\lambda =-\frac{1}{2}(2B-\dot\pi_\varphi -
\dot\pi_\vartheta ),\nonumber\\
 s_w\vartheta &=& -\frac{1}{2}(\dot B +\pi_\varphi +\pi_\vartheta ),\ \ 
s_w \pi_\varphi = 0,\ \ \ s_w \pi_\vartheta =0,\nonumber\\ 
 s_w u &=&0,\ \ \ s_w p_u =0, \ \ \ s_w c = 0,\ \ \ s_w B =0, \nonumber\\
 s_w \bar c &=&0.
\end{eqnarray}
However the symmetry operator $s_{\bar w}$ is not  an independent bosonic
symmetry transformation as shown below
\begin{eqnarray}
s_{\bar w}\varphi &=& \frac{1}{2}(\dot B +\pi_\varphi +\pi_\vartheta ),
\ \ \ s_{\bar w}\lambda =\frac{1}{2}(2B-\dot\pi_\varphi -
\dot\pi_\vartheta ),\nonumber\\
 s_{\bar w}\vartheta &=& \frac{1}{2}(\dot B +\pi_\varphi +\pi_\vartheta ),\ \ 
s_{\bar w} \pi_\varphi = 0,\ \ \ s_{\bar w} u =0, \nonumber\\
s_{\bar w} \pi_\vartheta &=&0,\ \ \
s_{\bar w} p_u =0, \ \ \ s_{\bar w} c = 0,\ \ \ s_{\bar w} B =0, \nonumber\\
 s_{\bar w} \bar c &=&0.
\end{eqnarray}
Now, it is easy to see the operators $s_w$ and $s_{\bar w}$ satisfy the relation
 $s_w +s_{\bar w} =0$. This implies, from Eq. (\ref{bos}), that
\begin{equation}
\{ s_b, s_d\} =s_w=-\{ s_{ab}, s_{ad}\},
\end{equation}
It is clear from the above algebra that the operator $s_w$ is the analogue of the
Laplacian operator in the language of differential geometry and the
 conserved charge for the above symmetry transformation is calculated as
\begin{equation}
Q_w=-i\left[B^2 +\frac{1}{2}(\pi_\varphi +\pi_\vartheta)^2 \right].
\end{equation}
Using the equation of motion, it can readily be checked that
\begin{equation}
\frac{d Q_w}{dt}=-i\int dx[2B\dot B+(\pi_\varphi +\pi_\vartheta)(\dot\pi_\varphi +
\dot\pi_\vartheta)]
=0.
\end{equation}
Hence $Q_w$ is the constant of motion for this theory.
\subsection{Ghost and discrete symmetries} 
Now we would like to mention  yet another symmetry of the system namely the ghost symmetry.
The ghost number of the ghost and anti-ghost fields  are 1 and -1 respectively,
the rest variables in the action
of the this theory have ghost number zero. 
Keeping this fact in mind we can introduce a scale transformation of the ghost field, under 
which the effective action is invariant, as
\begin{eqnarray}
\varphi \rightarrow \varphi, \vartheta\rightarrow \vartheta, \pi_\varphi\rightarrow 
\pi_\varphi, \pi_\vartheta\rightarrow 
\pi_\vartheta, u\rightarrow u\nonumber\\
 p_u \rightarrow p_u,\lambda\rightarrow \lambda, B\rightarrow B, c
\rightarrow e^{\Lambda}c,  \bar c
\rightarrow e^{-\Lambda}\bar c
\end{eqnarray}
where $\Lambda$ is a global scale parameter.
The infinitesimal version of the ghost scale transformation can be written as
\begin{eqnarray}
s_g\varphi &=& 0, \ \ s_g\vartheta = 0,
\ \ \ s_g\lambda =0,\nonumber\\
s_g \pi_\varphi &=& 0,\ \ \ s_g u =0, \ \ \ s_g \pi_\vartheta =0,\nonumber\\
s_g p_u &=&0, \ \ \ s_g c = c,\ \ \ s_g B =0, \ \ \ s_g \bar c =-\bar c.
\end{eqnarray}
The Noether conserved charge for above symmetry transformations is
\begin{equation}
Q_g =i[\dot{\bar c}c+\dot c\bar c]
\end{equation}
In addition to the above continuous symmetry transformation, the ghost
sector respects the following discrete symmetry transformations
\begin{equation}
c \rightarrow  \pm  i \bar c,\ \ \ \bar c \rightarrow  \pm  i c.
\end{equation}
The above discrete symmetry transformation is useful in enabling us to obtain
the anti BRST symmetry transformations from the BRST symmetry transformation and vice-versa.

\section{Geometrical cohomology} 
In this section we study the de Rham 
cohomological operators and their realization in terms of
 conserved charges which generate the symmetries for the
theory of self-dual chiral boson. 
In particular we point out the similarity between the algebra obeyed by de Rham 
cohomological operators and that by different BRST conserved charges.
\subsection{Hodge-de Rham decomposition theorem and differential operators}
The de Rham cohomological operators in
differential geometry obey the following algebra
\begin{eqnarray}
 d^2&=&\delta^2 =0, \ \ \Delta =(d +\delta )^2 =d\delta +\delta d\equiv \{d,\delta\}
\nonumber\\
\ \ [\Delta, \delta ]&=&0, \ \ [\Delta,d ]=0, \label{alg}
\end{eqnarray}
where $d, \delta$ and $\Delta$ are exterior, co-exterior and Laplace-Beltrami operator 
respectively.
The operators $d$ and $\delta$ are adjoint or dual to each other and  $\Delta$ 
is self-adjoint operator \cite{brac}.
It is well-known that the exterior derivative raises the degree of  form by
one when it operates on it (i.e. $df_n\sim f_{n+1}$). On the other hand, the dual-exterior 
derivative lowers the degree of a form by one when it operates on forms 
(i.e. $\delta f_n\sim f_{n-1}$). However $\Delta$ does not change the degree of form 
(i.e. $df_n\sim f_n$).
Where $f_n$ denotes an arbitrary n-form  object.

Let $M$ be a compact, orientable Riemannian manifold, then an inner product on the vector 
space 
$E^n (M)$ of $n$-forms on $M$ can be defined as \cite{gold}
\begin{equation}
(\alpha, \beta ) =\int_M \alpha\wedge \ast \beta,
\end{equation}
for $\alpha, \beta\in E^n (M)$ and $\ast $ represents the Hodge duality operator \cite{mor}.
Suppose that $\alpha$ and $\beta$ are forms of degree $n$ and $n+1$ respectively,
then following relation for inner product will be satisfied
\begin{equation}
(d\alpha, \beta )=(\alpha, \delta\beta ).
\end{equation}
Similarly, if $\beta$ is form of degree $n-1$, then we have the relation 
$(\alpha, d\beta )=(\delta\alpha, \beta )$. Thus the necessary and 
sufficient condition for $\alpha$ 
to be closed is that it should be orthogonal to all co-exact forms of degree $n$.
The form $\omega\in E^n (M)$ is called harmonic if $\Delta \omega =0$. Now let $\beta$ be a 
$n$-form  on $M$ and if there exists another $n$-form $\alpha$ such that $\Delta\alpha=\beta$,
then for a harmonic form $\gamma\in H^n$, 
\begin{equation}
(\beta, \gamma)=(\Delta\alpha, \gamma)=(\alpha, \Delta\gamma)=0,\label{Del}
\end{equation}
where $H^n(M)$ denote the subspace of $E^n(M)$ of harmonic forms on $M$.
Therefore, if a form $\alpha$ exist with the property that $\Delta\alpha =\beta$,
then Eq. (\ref{Del}) is necessary and sufficient condition for $\beta$ to be orthogonal
 to the subspace $H^n$.
This reasoning lead to the idea that $E^n(M)$ can be partitioned into three 
distinct subspaces $\Lambda^n_d$, $\Lambda^n_\delta$ and $H^n$ which are consistent
with exact, co-exact  and harmonic forms respectively.
Therefore, the Hodge-de Rham decomposition theorem can be stated as \cite{wan}: 

{\textit{ A regular differential form of degree n ($\alpha$) may be uniquely decomposed into 
a 
sum of  the harmonic form ($\alpha_H$), exact form ($\alpha_d$) and co-exact form 
($\alpha_\delta$) i.e.
\begin{equation}
\alpha =\alpha_H +\alpha_d +\alpha_\delta,
\end{equation}
where $\alpha_H\in H^n, \alpha_\delta\in \Lambda^n_\delta$ and $\alpha_d\in \Lambda^n_d$.}}
\subsection{Hodge-de Rham decomposition theorem and conserved charges} 
Using the canonical brackets given in Eq. (\ref{brac}) the conserved charges of all the 
symmetry transformation
can be shown to satisfy the following algebra 
\begin{eqnarray}
Q_{b}^2&=&0,\ \ Q_{ab}^2=0,\ \ Q_{d}^2=0, \ \ Q_{ad}^2=0,\nonumber\\
 \{ Q_b, Q_{ab}\} &=&0,\ \ \{ Q_d, Q_{ad}\} =0,\ \{ Q_b, Q_{ad}\} =0, 
\nonumber\\
\{ Q_d,Q_{ab}\} &=&0,\ [ Q_g, Q_b] = Q_b,\ \ [ Q_g, Q_{ad}]= Q_{ad},\nonumber\\
 \left[ Q_g, Q_d\right]&=&- Q_d,  \ \ [ Q_g, Q_{ab}]= -Q_{ab},\nonumber\\
 \{Q_b, Q_d\} &=&  Q_w = - \{ Q_{ad}, Q_{ab}\},\ \left[ Q_w, Q_r\right] = 0, 
\label{cgs}
\end{eqnarray}
where $Q_r$ generically represents the charges for BRST symmetry ($Q_b$), anti-BRST 
symmetry ($Q_{ab}$), dual-BRST symmetry ($Q_d$), anti dual-BRST symmetry ($Q_{ad}$) and
ghost symmetry ($Q_g$). We note that the 
relations between the conserved charges $Q_b$ and $Q_d$ as well 
as $Q_{ab}$ and $Q_{ad}$  mentioned in the last line of (\ref{cgs}) can be established  
by using equation of motions only.  

This algebra is reminiscent of the algebra satisfied by the de Rham cohomological operators
of differential geometry given in Eq. (\ref{alg}). Comparing (\ref{alg}) and (\ref{cgs}) we 
obtain following analogies   
\begin{eqnarray}
(Q_b, Q_{ad})\rightarrow d, \ \ (Q_d, Q_{ab})\rightarrow \delta, \ \ Q_w\rightarrow \Delta.
\end{eqnarray}
Let $n$ be the ghost number associated with a particular state $\left|\psi\right>_n$ 
defined in the total Hilbert space of states, i.e.,
\begin{equation}
iQ_g\left|\psi\right>_n = n\left|\psi\right>_n
\end{equation}
Then it is easy to verify the following relations 
\begin{eqnarray}
 Q_g Q_b\left|\psi\right>_n &=& (n+1)Q_b\left|\psi\right>_n,\nonumber\\
 Q_g Q_{ad}\left|\psi\right>_n &=& (n+1)Q_{ad}\left|\psi\right>_n,\nonumber\\
 Q_g Q_d\left|\psi\right>_n &=& (n-1)Q_b\left|\psi\right>_n,\nonumber\\
 Q_g Q_{ab}\left|\psi\right>_n &=& (n-1)Q_b\left|\psi\right>_n,\nonumber\\
 Q_g Q_w\left|\psi\right>_n &=& nQ_w\left|\psi\right>_n,\label{gh}
\end{eqnarray}
which imply that the ghost numbers of the states 
$Q_b\left|\psi\right>_n$, $Q_d\left|\psi\right>_n $ and $Q_w\left|\psi\right>_n $
are $(n + 1), (n - 1)$ and $n $ respectively.
The states $ Q_{ab}\left|\psi\right>_n$ and $ Q_{ad}\left|\psi\right>_n $
have ghost numbers $ (n - 1)$ and $(n + 1)$ respectively. 
The properties of $d$ and $\delta$ are mimicked by sets ($Q_b,Q_{ad}$) and ($Q_d,Q_{ab}$), 
respectively. It is evident from Eq. (\ref{gh}) that  the set ($Q_b,Q_{ad}$) raises 
the ghost number of a
state by one and on the other hand the set ($Q_d, Q_{ab}$) lowers the ghost number of 
the same state
by one.
 These observations, keeping the analogy with the Hodge-de Rham 
decomposition theorem, enable us to express any arbitrary state 
$\left|\psi\right>_n$ in terms of the sets
 ($Q_b, Q_d, Q_w$) and ($Q_{ad}, Q_{ab}, Q_w$) as
\begin{equation}
\left|\psi\right>_n = \left|w\right>_n +Q_b\left|\chi\right>_{n-1}+Q_d\left|\phi\right>_
{n+1},
\end{equation}
\begin{equation}
\left|\psi\right>_n = \left|w\right>_n +Q_{ad}\left|\chi\right>_{n-1}+Q_{ab}
\left|\phi\right>_
{n+1},
\end{equation}
where the most symmetric state is the harmonic state $\left|w\right>_n$ that satisfies 
\begin{eqnarray}
Q_w\left|w\right>_n &=&0, \ \ Q_{b}\left|w\right>_n =0, \ \ Q_{d}
\left|w\right>_n=0,\nonumber\\
Q_{ab}\left|w\right>_n &=&0,\ \ Q_{ad}\left|w\right>_n =0,
\end{eqnarray}
analogous to the Eq. (\ref{Del}).
It is quite interesting to point out that the physicality criteria of  all the fermionic 
charges $Q_{b},  Q_{ab}, Q_{d}$ and $ Q_{ad}$,
i.e.,
\begin{eqnarray}
Q_{b}\left|phys\right> &=&0, \ \ Q_{ab}\left|phys\right> =0,\nonumber \\ 
Q_{d}\left|phys\right> &=&0, \ \ Q_{ad}\left|phys\right> =0,
\end{eqnarray}
leads to the following conditions
\begin{eqnarray}
&&\pi_\lambda\left|phys\right> =0,\nonumber\\
&&(\pi_\varphi +\pi_\vartheta )\left|phys\right> =0.
\end{eqnarray}
That is the operator form
of the first-class constraint which annihilates the physical state as a consequence of
the above physicality criteria, which is consistent with the Dirac's
method of quantization of a system with first-class constraints.
\section{Conclusion} 
 BFV technique plays important role in the gauge theory to analyze the 
 constraint and the symmetry of the system. We have considered such a
powerful technique to study the theory of self-dual chiral boson. In particular we have 
derived the nilpotent BRST symmetry transformations for this theory using BFV technique.
Further we have studied the dual-BRST transformation which is also the symmetry 
of the effective
action and leaves gauge-fixing part of effective action invariant separately. 
Interchanging the role of ghost and antighost fields the anti BRST and anti dual-BRST 
symmetry  transformations are also constructed.
We have shown that the nilpotent BRST and anti dual-BRST symmetry transformations are 
analogous to the exterior derivative as the ghost number of  the state
 $\left|\psi\right>_n$ on the 
total Hilbert space is  increased by one when these operate on $\left|\psi\right>_n$
and the algebra followed by these are same as the algebra obeyed by the
de Rham cohomological operators.
 In the similar
 fashion the dual-BRST and anti BRST symmetry transformations are
 linked to the co-exterior derivative.
 The anticommutator of either the BRST and the dual-BRST transformations or
 anti BRST and anti dual-BRST transformations
 leads to a bosonic symmetry  which turns out to be
the analogue of the Laplacian operator. Further, the effective theory has a non-nilpotent
 ghost symmetry transformation which leaves the  ghost terms of the effective 
action invariant independently. Further we have noted that the   
 Hodge duality operator $(\ast )$ does not exist for the theory of self-dual chiral boson 
in  (1+1) 
dimension
because effectively this theory reduces to a theory in (0+1) dimension
due to self duality condition of fields ($\dot\varphi=\varphi'$ as well as 
$\dot\vartheta=\vartheta'$). 
 The algebra satisfied by the conserved charges is exactly same in appearance as the 
algebra of the de Rham cohomological operators of differential geometry. Thus we have 
realized that the theory for self-dual
chiral boson is a Hodge theory.

\vspace{.2in}
{\bf Acknowledgments}\\
\vspace{.07in}

{We thankfully acknowledge the financial support from the Department of Science and Technology 
(DST), Government of India, under the SERC project sanction grant No. SR/S2/HEP-29/2007.
We are thankful to the referee for his/her constructive suggestions and comments.}

\end{document}